\documentclass{article}

\PassOptionsToPackage{numbers, sort, compress}{natbib}
 \usepackage[dblblindworkshop, final]{neurips_2025}

\usepackage[utf8]{inputenc} % allow utf-8 input
\usepackage[T1]{fontenc}    % use 8-bit T1 fonts
\usepackage{hyperref}       % hyperlinks
\usepackage{url}            % simple URL typesetting
\usepackage{booktabs}       % professional-quality tables
\usepackage{amsfonts}       % blackboard math symbols
\usepackage{nicefrac}       % compact symbols for 1/2, etc.
\usepackage{microtype}      % microtypography
\usepackage{xcolor}         % colors
\usepackage{graphicx}       % RR: For images

\usepackage{hyperref}
\hypersetup{
    colorlinks=true,
    linkcolor=blue,
    citecolor=teal, % ForestGreen, MidnightBlue, Magenta
    filecolor=green,
    urlcolor=blue}

\title{Reversing the Lens: Using Explainable AI to Understand Human Expertise}
\workshoptitle{Machine Learning for Physical Sciences (ML4PS)}
\author{%
  Roussel Rahman\\
  Department of Photon Science\\
  SLAC National Accelerator Laboratory\\
  Menlo Park, CA 94025 \\
  \texttt{roussel.rahman@gmail.com} \\
  \And
  Aashwin Ananda Mishra\\
  Department of Machine Learning\\
  SLAC National Accelerator Laboratory\\
  Menlo Park, CA 94025 \\
  \texttt{aashwin@stanford.edu} \\
  \And
  Wan-Lin Hu\\
  Department of Photon Science\\
  SLAC National Accelerator Laboratory\\
  Menlo Park, CA 94025 \\
  \texttt{wanlinhu@gmail.com} \\
}

\begin{document}

\maketitle

\begin{abstract}
    Both humans and machine learning models learn from experience, particularly in safety- and reliability-critical domains. While psychology seeks to understand human cognition, the field of Explainable AI (XAI) develops methods to interpret machine learning models. This study bridges these domains by applying computational tools from XAI to analyze human learning. We modeled human behavior during a complex real-world task -- tuning a particle accelerator -- by constructing graphs of operator subtasks. Applying techniques such as community detection and hierarchical clustering to archival operator data, we reveal how operators decompose the problem into simpler components and how these problem-solving structures evolve with expertise. Our findings illuminate how humans develop efficient strategies in the absence of globally optimal solutions, and demonstrate the utility of XAI-based methods for quantitatively studying human cognition. 
    %Our findings illuminate how humans develop effective, near-optimal strategies in the absence of global solutions, and demonstrate the utility of XAI-based methods for quantitatively studying human cognition.
    
\end{abstract}

% Keywords: Explainable AI (XAI), Human Expertise, Bounded Rationality, Human-AI Alignment, Complex Reasoning
\section{Introduction}

As the horizon broadens for the application of machine learning (ML) and large language models (LLMs) in physical sciences, the importance of understanding AI reasoning in complex environments is at its highest. Real-world, physical environments are inherently complex and pose formidable challenges for both human and artificial agents. To ensure safe and reliable deployment of AI systems in such environments, we must first understand and then improve their reasoning capabilities. However, current benchmarks of reasoning often fail to reflect the complexity of real-world tasks, and detailed explanations remain limited to relatively simple tasks \citep{wang2022interpretability, bereska2024mechanistic, ying2025benchmarking}. Moreover, AI reasoning remains fragmented across domains (e.g., mathematical, spatial, or general reasoning), which hinders an integrated understanding.

We believe a general set of methods to model and investigate reasoning in complex environments can help greatly. Despite immense progress in AI reasoning, humans consistently outperform AIs in complex environments \citep[e.g.,][]{gigerenzer2008heuristics, bossaerts2017computational, gigerenzer2020bounded, rahman2022dynamics, sibert2025need}. This superiority stems from their ability to reason efficiently in solving problems of enormous complexity with limited computational resources. General representations of complex reasoning can help learn from human solutions in improving AI reasoning, as well as improving human-AI alignment. 

A rich line of research on human reasoning indicates that complex reasoning must be investigated and explained through processes or methods that are feasible to implement, given the environmental complexity \citep{newell1958elements,simon1962architecture, simon1971human, anzai1979theory, simon1998we, gigerenzer2020explain, rahman_gray2020topics, rahman2021precisemeasures}. Graph theory provides a general framework for achieving such process-level explanations of reasoning for both human and AI agents.
Here, we implement a set of graph-based methods, commonly used in ML and XAI, to model how humans solve a complex real-world task at various levels of experience. The experimental task we use is tuning a particle accelerator, which requires complex reasoning in a large and uncertain search space. We represent the whole task as weighted graphs of its parameters for three experience groups. Thereafter, we examine (1) the processes as subsets of task parameters using community detection algorithms and (1) the organization of the task parameters through hierarchical clustering. We find that the operators divide the task parameters into three subsets regardless of their experience level. However, we also find fine-grained changes in the structure underneath the similarity of task partitions.

\section{Modeling and Explaining Behavior in Complex Environments}

ML has been applied to a wide range of physical sciences, such as statistical physics, particle and quantum physics, quantum computing, and chemistry \citep{carleo2019machine}. As an example relevant to our experimental paradigm, in particle accelerator operations, ML methods enable simulations of control systems, anomaly detection, uncertainty quantification, system design, and active control \citep{edelen2018opportunities, edelen2019machine, mishra2021uncertainty, gupta2021improving, duris2020bayesian}. 

Importantly, we need to explain AI reasoning not just in terms of inputs and outputs, but also in terms of the formation of higher-level concepts necessary for efficient problem-solving \citep{olah2020zoom, nanda2023progress, bereska2024mechanistic}. The problems in physical sciences are generally complex and uncertain, which eliminates the possibility of finding optimal solutions using the traditional views of rationality \citep{simon1962architecture, simon1976substantive}. For such problems, humans use \textit{bounded rationality}; that is, they approximate good enough solutions using heuristics that frequently outperform the state-of-the-art optimization algorithms in complex environments \citep{simon1955behavioral, gigerenzer2008heuristics, gigerenzer2020bounded}. Crucially, the ML models are not immune to complexity; thus, to improve their performance in complex environments, we need to teach them to reason efficiently as humans do.

% -- Humans navigate mental spaces in a manner similar to navigating physical environments: 
We believe graph models of complex behavior provide a promising path to general explanations. Graph-theoretic models serve as the foundation for cognitive network science, which has been exceptionally successful in explaining complex problem solving and reasoning of humans \citep{siew2019cognitive, kenett2020cognitive}. Graph-based methods also serve as a bedrock for improving and explaining the performance of neural networks \citep{graphbook, zhang2022graph, bereska2024mechanistic}. In this work, we use graphs to model human performance in the complex task of tuning a particle accelerator and demonstrate the efficacy of graph-theoretic measures in capturing how humans navigate and master the task.

\section{Methods Used}

\subsection{Experimental Task: Tuning Particle Accelerators}

The accelerator we use is the Linac Coherent Light Source, a Free-Electron LASER (FEL) at SLAC National Accelerator Laboratory. The goal of FEL tuning is to maximize the pulse intensity of the resultant X-ray beams, using a set of 27 tuning parameters. The search space of parameter values is enormous, making the task extremely complex. For illustration, there are $27! \approx 1.09\times10^{28}$ possible sequences to adjust the parameters in and $\approx 5.45 \times 10^{20}$ ways to partition the set into subsets.

\subsection{Dataset and Participants}

To examine how the operators handle this complexity, we use a large archive of about 350,000 texts they logged during operations between 2009 and 2022. We obtained Institutional Review Board (IRB) approval to use this dataset for our study, and detailed measures were implemented to anonymize the data. To extract information related to FEL tuning and identify the parameters used, we parsed the logs using a host of natural language processing methods. For details of the steps, please see \citep{rahman2024network}. The resulting data were divided into three groups based on experience level: (1) Novices (<1 year of experience), (2) Intermediates (1-4 years), and Experts (>4 years).

\subsection{Graph Construction and Analysis}

For each group, the graphs were constructed using the 27 parameters as nodes and the co-occurrences between parameters as edge weights. Thereafter, we examined the presence of groups (using community detection) and the organization of the parameters (using hierarchical clustering).

\subsubsection{Community Detection}
Communities are local structures in graphs, consisting of a subset of nodes with high edge density within the subset and low edge density elsewhere. As community detection is an NP-hard problem, finding optimal partitions is intractable beyond small graphs, and heuristic-based approaches are used for large graphs \citep{fortunato2016community}. We used two popular algorithms for community detection: (1) the Louvain algorithm, and (2) spectral clustering. The strength of partitions is measured by modularity, which compares the actual density of edges within communities to the density expected at random \citep{newman2004analysis}. Modularity ranges between $[-1, 1]$. Values close to 0 indicate partitions no better than random, and values of 1 represent perfectly separated partitions. Values between 0.3-0.7 are considered to indicate \textit{strong} partitions \citep{newman2004finding, newman2004analysis}.

\subsubsection{Hierarchical Clustering}
Communities represent sets of nodes that cluster together, but do not reveal the structure of the nodes; for this purpose, hierarchical clustering is a widely used method. We use agglomerative hierarchical clustering based on linkage methods \citep{hastie2009elements, james2013introduction}; that is, we begin with individual nodes at the lowest level and cluster nodes based on pairwise distances as we progress to increasingly higher levels, until the cluster encompasses all nodes and converges to the entire graph. The distances are computed using complete linkage, which averages the pairwise distances between all nodes in two clusters.

\section{Results}
\subsection{Consistent Communities in Graphs across levels of experience}

\begin{figure}[!b]
    \includegraphics[width=1.0\textwidth]{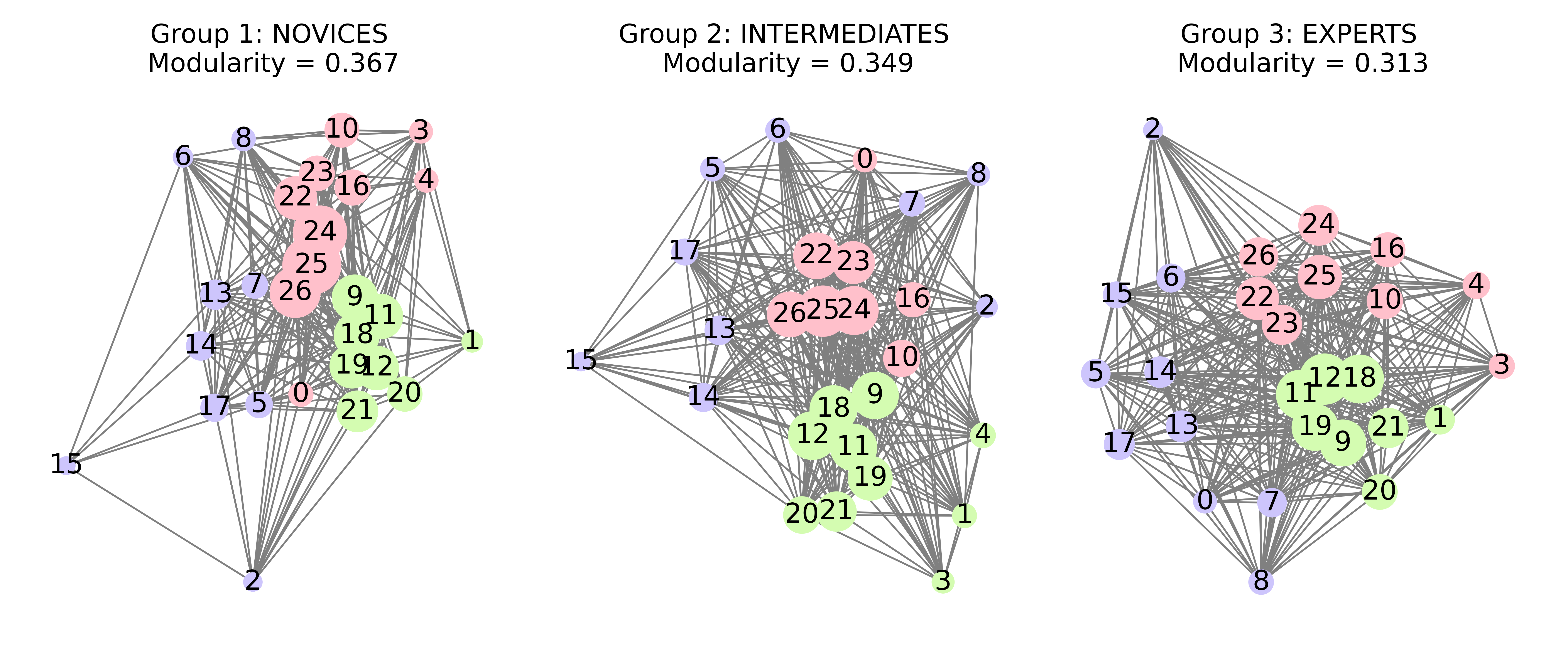}
    \caption{Graphs of FEL tuning for three groups of operators. The node sizes represent the PageRank values, and the distances between nodes represent the edge weights. Communities (denoted by colors) were identified using the Louvain algorithm and verified using spectral clustering. }
    \label{fig:three_groups_whole_networks}
\end{figure}

Figure \ref{fig:three_groups_whole_networks} displays the graphs for three groups of operators, where the nodes are colored according to the communities detected by the Louvain algorithm and verified using spectral clustering. Modularity values for the three groups (0.367, 0.349, and 0.313, respectively) are all above 0.30, indicating strong partitions.
%For all three groups, modularity values of the partitions are well above 0.30 (0.367, 0.349, and 0.313, respectively), suggesting strong partitions.
Importantly, the three groups demonstrate remarkable similarities in categorizing the subtasks into communities. We find exactly three communities in the networks for all groups. These communities are also largely similar, with only one or two subtasks being classified differently across groups (e.g., Parameter 0 for the experts and Parameters 3 \& 4 for the intermediates). Upon consulting with domain experts, we learned that the community denoted in green consists of parameters related to beam transport and steering, the purple community corresponds to parameters that affect beam energy and compression, and the pink set consists of all other parameters. 

%In the rest of this article, we will denote the three communities respectively as (1) the beam transport (green) community, (2) the beam energy/compression (purple) community, and (3) the miscellaneous (pink) community.

The strong partitions indicate that humans divide the complex task into parts of manageable complexity. The similarities of communities demonstrated by the groups are quite striking, considering the extremely large number of possible partitions ($\approx 5.45 \times 10^{20}$). These similarities strongly suggest that operators at all stages of expertise can effectively recognize and categorize parameters into similar groups. Therefore, any differences in tuning performance with expertise are unlikely to stem from improvements in categorizing different parameters into communities.

\subsection{Evolving Hierarchical Structures within Communities with Experience}

\begin{figure}[!t]
    \centering
    \includegraphics[width=\textwidth]{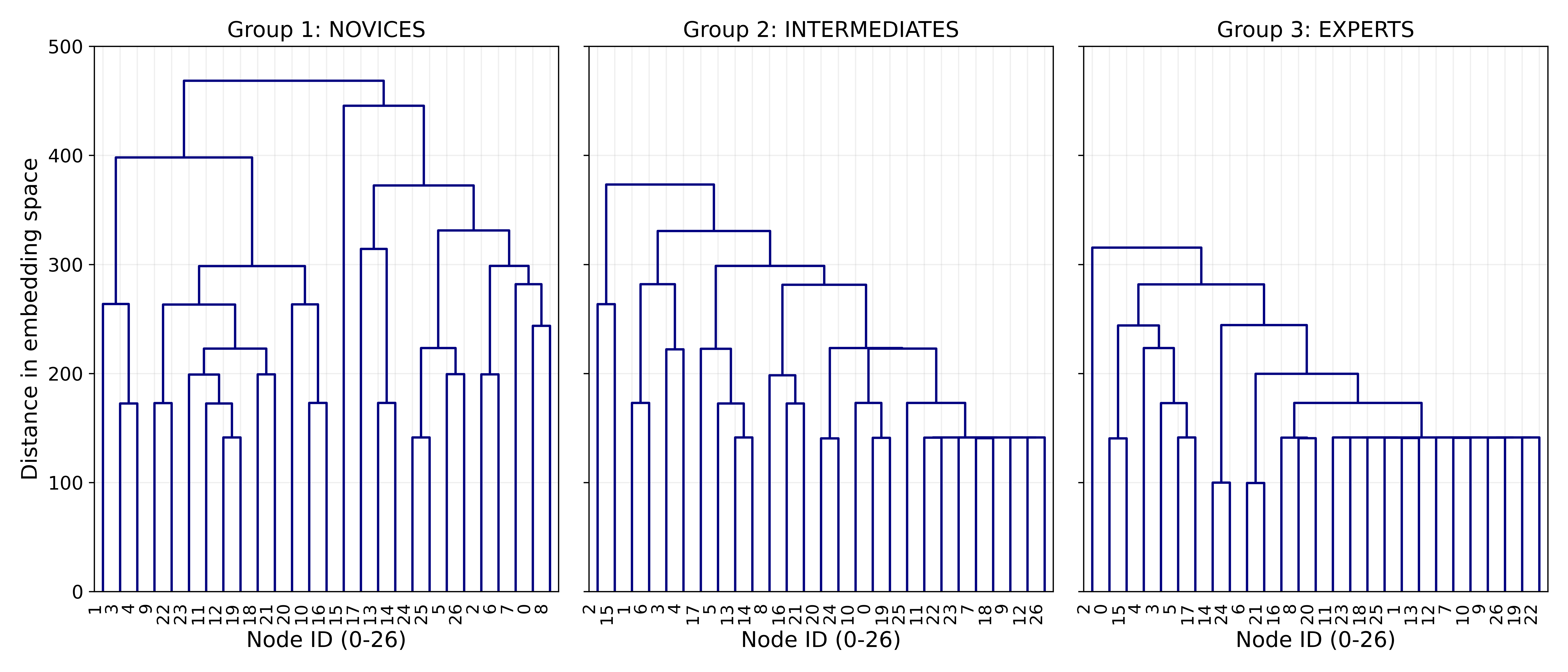}
    \caption{Hierarchical clustering of FEL tuning graphs for three experience groups. The height shows the distance between elements in the embedding space. The elements are clustered together based on this distance.}
    \label{fig:three_groups_hierarchy}
\end{figure}

% Though subtasks are consistently grouped into these three broad communities, there are differences in how subtasks converge into communities among the three stages of operator expertise. 

To examine how subtasks are organized into communities, we performed hierarchical clustering based on linkage methods that aim to group elements based on their distances in graph embedding space. The results are shown as dendrograms in Figure \ref{fig:three_groups_hierarchy}. The height of the dendrograms represents the distance at which two nodes are considered to belong to the same cluster. The horizontal connections mark the distance at which the elements connect or become part of the same group. 

As we see, the dendrograms decrease in height with increased expertise, indicating that with experience, the subtasks became closer in distance and the graphs became denser. For novice operators, the subtasks are grouped together at much higher distances than for the other groups. Moreover, the distances varied considerably more for the novices than for others. Finally, the differences in structure appear to be larger between the novices and the intermediates than between the intermediates and the experts, reflecting a steep learning curve for novice operators.

\section{Conclusions}

Our two sets of results indicate that, underneath the similarities in the communities, the frequency and sequencing of subtasks may change considerably with expertise. Surprisingly, the communities remained the same at all experience levels, despite the large scope of differences. The modularity values also indicate \textit{strong} partitions that are unlikely to be found at random. These results strongly support a divide-and-conquer strategy often observed in human problem solving. As optimizing parts of complex systems does not guarantee global optimality, this strategy is a boundedly rational approach, one that enables us to solve problems of enormous complexity using limited computational resources. To improve AI reasoning in complex environments, we need to train models to be efficient in resource use, for which human performance provides a roadmap.

While our study needs to be replicated for a larger set of tasks to generalize the findings, it highlights the need to examine and explain AI reasoning with models that accurately reflect the complexity of the task at hand. As there are numerous paths of reasoning, we need to specify the actual processes used by the AI agents, as we do for human agents. Otherwise, we may expect abilities or processes beyond the agents for the given problem. Notably, in cognitive models, human process or strategy selection is often modeled and explained as rational meta-reasoning among alternatives based on some form of reinforcement learning \citep{lieder2017metareasoning, lieder2020resource, gonzalez2003instance, sun2008introduction, sibert2025need}, but at the cost of modeling a part of the process as a \textit{blackbox} \citep{gigerenzer2020explain}. Therefore, general methods to probe intelligent behavior and reasoning in complex environments may lead to an integrated understanding and accurate benchmarks, helping to maximize the effectiveness of human-AI teams in complex, uncertain environments of the real world.

\begin{ack}

We are especially grateful to Jane Shtalenkova for her immense contributions to this research -- her insights, dedication, and technical expertise were instrumental throughout. 
We would like to thank the accelerator operations team at SLAC National Laboratory, who generously allowed us into the control room and shared their wealth of system expertise to make our work possible. Specifically, we would like to thank: the operators for giving us insight into their unique and incredible skill set and generating the elog data set; Peter Schuh, Johnny Warren, and Alex Saad for supporting and enabling this research and connecting us with the operators; Benjamin Ripman and Janice Nelson for consulting on the design of this study with special thanks to Matt Gibbs who did all that and helped us navigate a myriad of networking and software development challenges. We would like to thank John Schmerge for being our upper-management advocate and extend our deepest gratitude to him and Mike Dunne for securing the additional funding to complete this research.

This research was supported in part by the SLAC Laboratory Directed Research and Development (LDRD) program under the project ``Performance Optimization for Human-in-the-Loop Complex Control Systems,'' led by Wan-Lin Hu.
\end{ack}

\section*{Ethics and Compliance Statement}
This study was reviewed and approved by the Institutional Review Board (IRB) at Stanford University. The secondary dataset used is a subset of the archival data at SLAC National Accelerator Laboratory, and all identifying information was anonymized prior to analysis.

\section*{Code and Data Availability}
A Python implementation of all methods used can be found here: {\small \url{https://github.com/Roussel006/Expertise-in-Operating-Particle-Accelerators-through-Network-Models-of-Performance}}. The dataset can be found here: \url{https://osf.io/qmt2x/}.

\bibliographystyle{unsrt}

\bibliography{bibliography}

\end{document}